# Laser-plasma accelerated protons: energy increase in gas-mixtures using high mass number atomic species

**Tadzio Levato [1,2,*], Leonardo V. Goncalves [1], Vincenzo Giannini [3]**

[1] Institute of Physics of the ASCR, ELI Beamlines Project, Na Slovance 2, 18221 Prague, Czech Republic;
[2] Intense Laser Irradiation Laboratory (ILIL), Istituto Nazionale di Ottica – Consiglio Nazionale delle Ricerche (INO-CNR), Via Moruzzi, 1, 56124 Pisa, Italy;
[3] Instituto de Estructura de la Materia (IEM), Consejo Superior de Investigaciones Científicas (CSIC), Serrano 121, 28006 Madrid, Spain;

* Correspondence: tadzio.levato@eli-beams.eu; tadzio.levato@gmail.com;

**Abstract**: The idea of using a gas-mixture comprising atoms with high mass number in order to increase proton energies in laser induced plasma acceleration at under critical density is investigated by means of 2D PIC (Particle-In-Cell) simulations. Comparing and discussing the case of a pure hydrogen plasma, and that of a plasma containing higher mass number species with a small percentage of hydrogen, we demonstrated that the mixture enhances the energies of the accelerated protons. We also show that using a gas-mixture introduces the possibility of using the densities ratio to change the relative acceleration of the species.

## 1. Introduction

After the advent of the Chirped-Pulse-Amplification (CPA) method [1 - Gerard] the growing interest in high power femtosecond laser systems has steered various concepts aiming at using laser-matter interaction for particle acceleration. The use of the so-called Laser Wake-Field Acceleration (LWFA) technique [2 - Tajima] has led to the production of about 8 GeV electrons over very short distances of roughly 10 cm [3 - Wim] in under-critical plasma using gas-filled capillaries. The use of the entitled TNSA (Target Normal Sheath Acceleration) has generated up to 100 MeV proton bunches in over-critical plasmas using solid targets [4 - Higginson et al. 2018].

The intermediate range of near-critical density (NCD) plasmas has also been experimentally used to accelerate ions at the plasma output interface (refer to Figure 1 and caption for the description of "plasma output" and "input"). Typically, such densities of $10^{19-21}$ e$^-$/cm$^3$ were reached by pre-illuminating a solid target to generate an expanding pre-plasma region [5 - Matsukado et al. 2003], or by operating precise high density gas-jet [6 - Willingale et al. 2006].

Over the last decades, laser proton acceleration was mainly performed in over-critical plasmas using solid targets that are understandably locally destroyed during the final phase of the process [7 - A. Macchi, M. Borghesi and M. Passoni 2013]. Using solid targets to work above the critical density results therefore in the introduction of a limiting factor in the repetition rate of the entire process, which is due to the need of refreshing the target surface after every single laser shot. The possibility arising from the use of a gas-jet, capable of operating at high repetition rate, or even in a continuous-flux mode, provides a significant contribution to overcome the limiting factor affecting the repetition rate of the process. Moreover, the use of high density gas-jet to accelerate protons at near under-critical density plasmas is also interesting from the view point of physics, since it offers an alternative method to the use of very high electric fields acting for a very short period of time that represents the approach of interaction with solid targets: the possibility of using moderate but longer lasting electric fields.

The discussion following [6] has pointed out that two main mechanisms [8 Bulanov and Esirkepov 2007 - 9 Willingale et al. 2007] can be identified to play a role in the proton acceleration following the interaction of an under-critical target with high power pulse:

i) one is the acceleration due to the electrostatic field related to space charge effects (SCE),
ii) the other one is the acceleration due to the induced electric field related to "magnetic vortex acceleration" (MVA).

In particular, it was shown [9] that acceleration due to the electrostatic field was 2/3 of the total accelerating field in [6]. Additional works have shown experimental results in near-critical over-dense plasmas [10 – Yogo et al. 2008] using very thin plasma. The near critical condition needed also investigation of the propagation properties of the high intensity laser in such a thin plasma [11 - Willingale et al. 2009] showing an increased efficiency in the proton acceleration mechanism for slightly under-critical plasma (0.9 $n_c$). Similar works [12 – Fukuda et al. 2009] confirmed the importance of the time varying fields at the plasma output also in the case of ions using gas-cluster target [13 – Faenov et al., 14 - Matsui et al.]. Such time varying fields has been shown to be enhanced by increasing the hot electron transport at high energies resulting in a fast advent of magnetic fields at the plasma output [15 – Willingale et al. 2010]. The use of particularly thin plasma slab obtained by collisionless plasma shock resulted in the acceleration of mono-energetic protons [16 – Haberberger et al. 2012], and in the identification of the density range needed to accelerate collimated ions [17 – Helle et al. 2016], also opening the possibility to increase the particle energy using a staged acceleration concept [18 – Ting et al. 2017] and improving the quality of the accelerated bunch tailoring the plasma profile [19 – Wan et al. 2019]. Other works have investigated experimentally, on solid targets, the use of mixed species plasma [20 – Pak et al. 2018, 21 - X. F. Shen et al. 2019], the role of nanostructure on the front-surface [22- Margarone et al. 2012] and the role of coating on the rear-surface [23 - Betti et al. 2009]. All the above mentioned studies are important to improve the efficiency of a laser-plasma proton accelerator.

The interest in a laser based proton accelerator cover different fields, from the most demanding hadron-therapy [24- CNAO, 25 – Cirrone et al. 2004] extensively studied in literature and requiring high-energy protons (250 MeV), to the proton-induced-xray-emission (PIXE) technique [26 – Barberio and Antici 2019] in which a proton bunch irradiates a sample to induce x-ray fluorescence [27 – Ryan 2001], the x-ray source is then imaged using dedicated pin-hole cameras techniques [28 – Labate et al 2012, 29 – Levato et al. 2010, 30 – Romano et al. 2016] able to image x-ray sources in extreme conditions as in laser-plasmas, where such techniques have allowed for the study of fast electron dynamics [31 – Zamponi et al. 2010] specifically thanks to the method adopted to "count" the single x-ray photons [32 – Levato et al. 2008, 33 – Labate et al. 2008]. The concept of all-optical radiation sources [34 – Gizzi et al. 2013] is widely used in laser-based under-dense plasma electron accelerator, since this is a great advantage for medium and small scale research laboratories [35 – Labate et al. 2016, 36 – Koester et al. 2015] offering the possibility to irradiate samples from industrial to biomedical interests.

In the present paper, we focus on proton acceleration mechanism in under-critical plasma regimes. Working with under-critical plasma is advantageous for a few practical aspects, the flexibility of the repetition rate as anticipated before, the simplicity in monitoring the interaction during the propagation [37 – Gizzi et al. 2011] and, as we will consider below in the text, for the flexibility to mix different atomic species having the possibility to arbitrary tune their relative densities. Experimental tests on proton acceleration have demonstrated also the possibility to use kHz laser offering an important advantage in term of compactness and versatility of the source [38 – Thoss et al. 2003, 39 - Morrison et al., 2018] using liquid targets. Different projects and research centers base their activity on the concept of plasma-acceleration, and we should mention in particular the case of ELI [40 – ELI White book] and some of the acceleration infrastructures within it [41 – ELIMAIA, 42 - HELL].

**2. Aim of this study and outline**

In this paper we examine the possibility of using a gas-mixture for near-under-critical proton acceleration in laser-plasma interaction, introducing heavier atoms than hydrogen in the background plasma. A similar idea, using solid-targets, has been proposed in [21]. Here we reconsider the two acceleration mechanism identified in [6,8-9] (points i) and ii) above) by investigating the role in proton acceleration played by a plasma mixture containing high mass number ions. In fact, since the heavier ions move slower than protons, they allow for the space charge effect at the plasma output to survive for longer times. This allows a longer acceleration time of the protons by the background plasma. The use of heavier atomic species in a gas-target introduces a new feature regarding the control of the acceleration of the lighter ions: i.e. the additional possibility of tuning the acceleration ratio by means of the density ratio of the species. This point will be discussed in the "modelling section" of this paper in which the results of 2D-PIC simulations, meant as a simplified "virtual experiment", are analyzed. The analysis will focus on the discussion of the two mechanisms identified above in a simplified setting which privileges a fluid description (see sections 5-6).

In particular, the first part of the analysis will describe the electrostatic acceleration ratio of the two species in a multi-fluid description: once the ultrarelativistic electrons escape from the plasma, the latter remains positively charged hence pushing the protons.

However, from the solid-target case [21] in which only a capacitor-like acceleration mechanism takes place, the gas-target case (under-critical case), as already described in [6-7-8-9], makes the two mechanism i) and ii) to coexists. This because the ultrarelativistic electron population rapidly generates a magnetic vortex structure at the plasma output, which induce an electric field that further accelerates the protons.

The second part of the modelling focuses on the acceleration ratio of the two species due to the induced fields.

## 3. Numerical models and simulation setup

The laser-plasma interaction is simulated by means of the 2D PIC code EPOCH [43 – Arber et al. 2015] using a basic scenario, in which two different atomic species are involved. The intent is to maintain the study as much as possible near to a realistic experimental case.

The idea of using different atomic species in the background plasma is considered in ideal conditions. In the present cases, a flat density plasma with a step-like transition from vacuum to plasma is used in combination with a decreasing density at the plasma output. The 2D PIC numerical box is of 220um long and 40um wide. The first 120um (plus the initial -20um) consists of a flat density plasma whereas the last 80um are used to decrease the density to zero by a half-Gaussian. This basic density profile has been used in order to put the emphasis at the plasma output where the proton acceleration takes place, this is obtained considering a plasma output profile more similar to the one associated to a real case rather than a simple ideal step-like transition. On the other hand, as the input plasma profile mainly affects the focusability of the laser pulse (whose study is beyond the scope of this article) an ideal step-like transition is therefore considered at the input plasma profile. The convolutional perfectly matched layer boundary conditions are used in all the 2D-PIC simulations (see reference [43] and references therein). All the simulations presented here are carried out with a spatial resolution of $\lambda/20$ in both directions, with 12 particles per cell, using a $\lambda=800$nm linearly polarized laser pulse to a peak intensity of $10^{20}$ W/cm². All the simulations performed are stopped after 1 ps from the beginning of the interaction (if not differently stated explicitly) giving only a partial indication of the final energy of the accelerated particles.

The idea is to study a common experimental case of a focal spot around 3 um (FWHM) using 1J-class, 30 femtosecond laser pulse as shown in the conceptual setup of **Figure 1**.

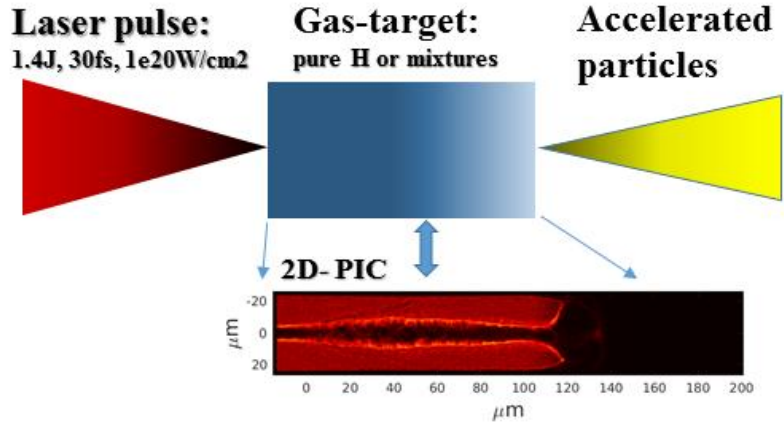

**Figure 1:** conceptual setup of the experiment simulated into the 2D-PIC. The laser enters at the left side which we will name hereafter "plasma input". The right edge will be hereafter named "plasma output".

The laser enters at the left side which we will name hereafter "plasma input". The right edge will be hereafter named "plasma output". The high intensity laser pulse pushes the electrons by means of the ponderomotive force during the propagation.

In order to verify the effect of the background plasma density, as well as to select the best case for the subsequent comparison with the plasma-mixture case, our starting point has been to consider pure hydrogen plasmas. We should note that despite the large number of different simulations that were performed operating at different gas-mixtures and that as matter of fact, in the whole, entirely confirmed the validity of the idea of using higher mass number ions to enhance proton acceleration, only the most relevant cases aiming towards that demonstration are shown. Additional investigation bearing in mind the output scale length plasma density and the optimal balance between the hydrogen fractions (%) in the gas mixture may contribute to further optimize the process.

Such optimal percentage will of course be altered for the case of ion acceleration (for example the He and C ions suitable for hadron-therapy)

## 4. Results of the 2D-PIC

*4.1 The case of pure hydrogen*

**Figure 2** shows the comparison of two snapshots of the laser plasma interaction at different plasma densities simulated by 2D PIC after 1000fs from the pulse entry (-20 um) in the simulation box (so that the laser pulse has already left the simulated model's area). The electrons (upper row) and protons (middle row) density distributions are shown for both pure hydrogen plasmas, respectively at a flat density of $7 \times 10^{19}$ e$^-$/cm$^3$ (left) and $2 \times 10^{20}$ e$^-$/cm$^3$ (right). The lower rows show the protons (black) and the electrons (green) energy versus the axis of the laser propagation. The comparison shows energetic protons in the same region of density decreasing for both cases, and an increase of the proton energy as expected for the higher plasma density case, up to about 5 MeV maximum (at 1000fs from the beginning of the interaction).

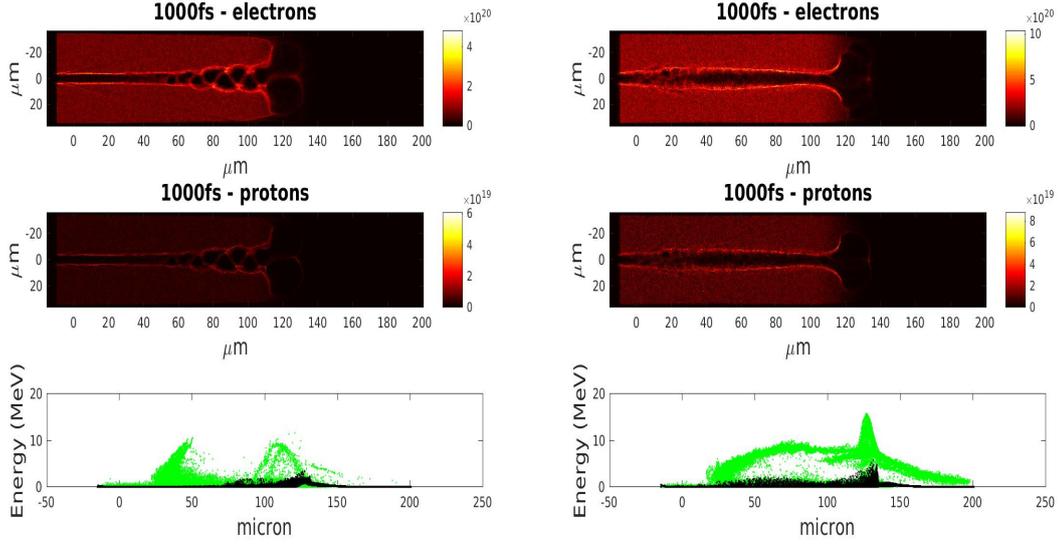

**Figure 2:** snapshots of the 2D PICs after 1000 fs from the laser pulse entry in the simulation box of a pure hydrogen plasma at a flat density of 7x10¹⁹ e⁻/cm³ **(left)**, and 2x10²⁰ e⁻/cm³ **(right)**. The electron (upper row) and proton (middle row) density distributions are shown for both cases together with the protons (black) and electrons (green) energy spectra versus the laser propagation direction (lower row).

As it is clearly shown, the proton energy increases in the region in which the plasma density starts to decline (the region in between 100 and 150 micron). We should mention however that in the present paper we are not interested in the study of the interdependence between the proton energy increase and the density decrease scale length. For this specific configuration the maximum proton energy, at 1000fs after the beginning of the interaction (so that the laser pulse has already left the simulated model's area) of the higher density case at 2x10²⁰ e⁻/cm³ approximately reaches a value of about 5 MeV, with the proton bunch longitudinally localized inside an interval of about 10um of the laser propagation axis. In particular, the proton energy increase from about 3 to 5 MeV respectively for the background plasma densities of 7x10¹⁹ and 2x10²⁰ e⁻/cm³ (at 1000fs from the beginning of the interaction). An increase in the background plasma density (still for the case in which a pure hydrogen plasma is considered) would lead to a further noticeable energy increase. A too high plasma density would prevent the laser propagation up to the "plasma output", inhibiting the entire proton acceleration process due to laser erosion.

**Figure 3** shows the snapshot of the laser plasma interaction simulated by 2D PIC, after 1000fs from the pulse entry in the simulation box, in the case of a pure hydrogen plasma at a flat density of 7x10²⁰ e⁻/cm³. As it is shown, in this case the laser pulse is not capable to propagate over few tens of microns at such plasma density inhibiting the entire proton acceleration process.

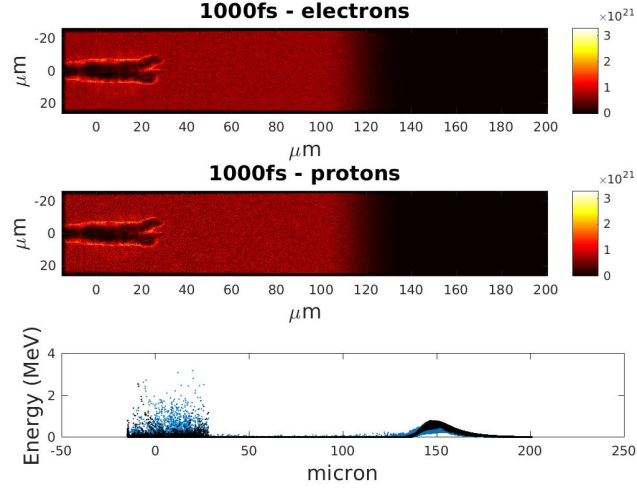

**Figure 3:** snapshot of the 2D PIC after 1000 fs from the pulse entry in the simulation box of a pure hydrogen plasma at a flat density 7x10$^{20}$ e$^-$/cm$^3$. The electron (upper row) and proton (middle row) density distributions are shown for both cases together with the protons (black) and electrons (blue) energy spectra versus the laser propagation direction (lower row).

*4.2 The case of hydrogen with a heavier species*

For the comparison of the effect of different mass numbers atoms in the background plasmas, we use the same plasma profile both in scale length and electron densities considered in the previous case.

Xenon (Xe), with A=131, stands out as a promising candidate in the selection process for a higher mass number noble gas (at room temperature and excluding the radioactive case of Radon) for the use of a gas-mixture-target. Nevertheless, also the cases of Kr, Ar, Ne and N could be considered for proton acceleration. The higher mass number case of Xe can be considered to accelerate heavier ions than protons such as He and C, both useful for hadron-therapy.

**Figure 4** shows the snapshots of the laser-plasma interaction for a fully-ionized 95% Xe - 5% H plasma mixture simulated by 2D PIC after 1000fs from the pulse entry (-20 um) in the simulation box (so that the laser pulse has already left the simulated model's area). For a simpler comparison with the case of Figure 1 (right), the total electron density is set to be **2x10$^{20}$ e$^-$/cm$^3$**, the plasma profile considered is the same. The electron (upper row) and proton (middle row) density distributions are shown with the protons (black) and electrons (green) spectra versus the laser propagation direction. The proton energy reaches approximately the 10 MeV threshold at the plasma output, confirming that the simple use of higher mass number ions promotes the acceleration process with an energy increase of about a factor 2 (at 1ps from the beginning of the interaction) when compared with the pure hydrogen case, still at the same electronic density and plasma profile.

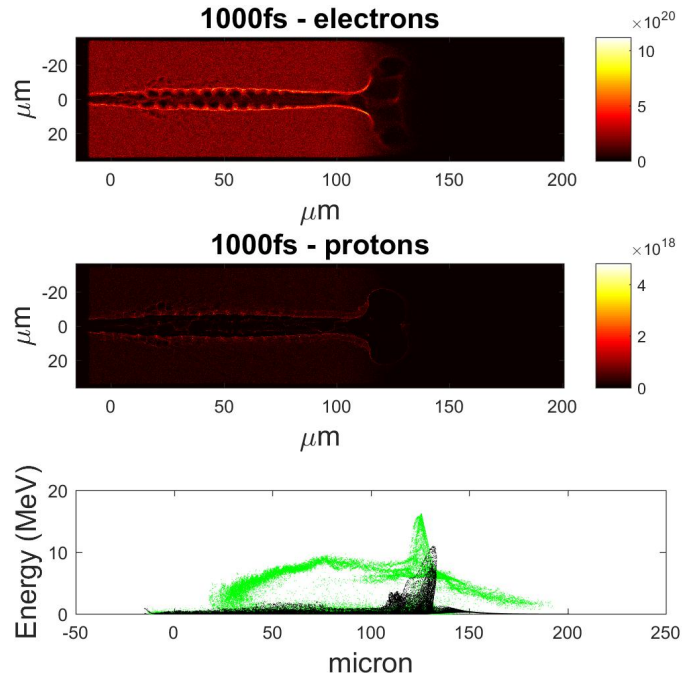

**Figure 4:** snapshots of the 2D PICs after 1000 fs from the laser pulse entry in the simulation box of a fully-ionized 95% Xe - 5% H plasma mixture at a flat density of $2 \times 10^{20}$ e$^-$/cm$^3$ for a comparison with the case of Figure 1-right. The electrons (upper row) and protons (middle row) density distributions are shown for both cases together with the protons (black) and electrons (green) energy spectra along the laser propagation direction (lower row). An energy increase of about a factor 2 is evident with respect to the pure hydrogen plasma case (**Figure 1**-right)

As anticipated, even lower mass number atoms and less rare gases than Xe can be used based on this idea of including different mass number atoms to promote the acceleration of the lighter one.
As an example we also consider Ar and N to accelerate protons.

**Figure 5** shows the snapshots of the laser-plasma interactions for a fully-ionized 95%**Ar** - 5%H (left) and 95%**N** - 5%H plasma mixture simulated by 2D PIC after 1000fs from the pulse entry (-20 um) in the simulation box (so that the laser pulse has already left the simulated model's area). The total electron density is set to be **$2 \times 10^{20}$ e$^-$/cm$^3$** for a simpler comparison with the case of Figure 1 (right), also the plasma profile considered is the same. The (upper row) and proton (middle row) density distributions are shown with the protons (black) and electrons (green) spectra versus the laser propagation direction.

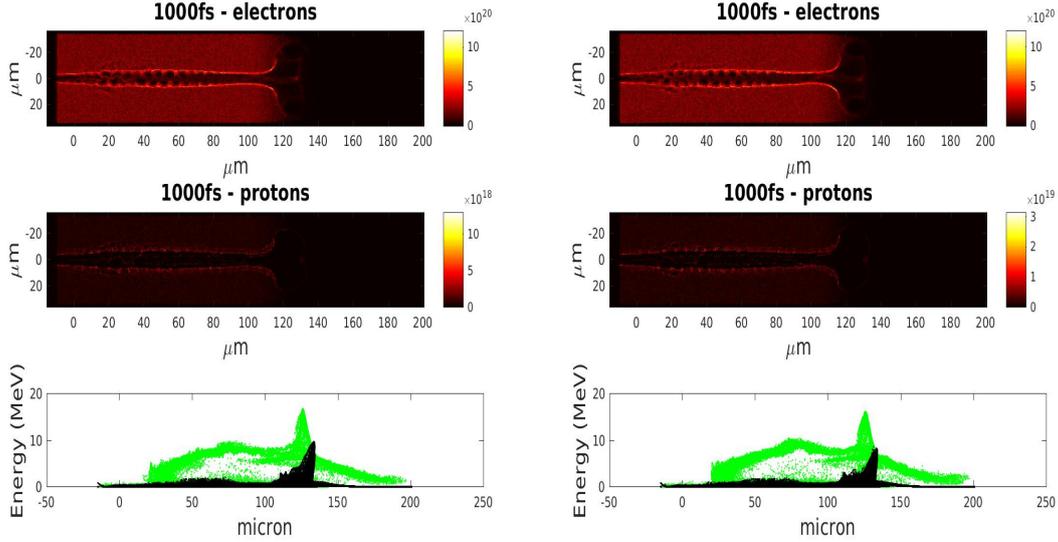

**Figure 5:** shows the snapshots of the laser-plasma interactions for a fully-ionized 95%**Ar** - 5%**H** (left) and 95%**N** - 5%**H** (right) plasma mixture simulated by 2D PIC after 1000fs from the pulse entry (-20 um) in the simulation box (so that the laser pulse has already left the simulated model's area). The total electron density is set to be **$2 \times 10^{20}$ e$^-$/cm$^3$** for a simpler comparison with the case of Figure 1 (right), also the plasma profile considered is the same. The (upper row) and proton (middle row) density distributions are shown with the protons (black) and electrons (green) spectra versus the laser propagation direction.

The proton energy reaches approximately the value of 10 MeV at the plasma output, which leads to an energy increase of about a factor 2 when compared with the pure hydrogen case, still at the same electronic density and plasma profile (at 1000fs from the beginning of the interaction).

Besides we should note here that in all the above mentioned 2D-PIC simulations the general condition was that of the presence of a proton density much lower than the considered heavier atom density.

## 5. Interpretation and discussion: acceleration ratio of two different species

To estimate the acceleration ratio of different plasma species we consider the two main mechanisms at play in the proton acceleration as discussed in [6,8,9], nonetheless following an idealized scheme that privileges a fluid-like description.

The first mechanism (indicated as i) above in the text) consists in the electrostatic field derived from the space charge distribution: since a bunch of electrons escapes from the plasma, the latter remains positively charged.

The second one (indicated as ii) above in the text) consists in the electric field induced by the magnetic vortex structure: the ultra-relativistic electrons escaping from the plasma generate a magnetic field that induces an electric field further pushing the protons.

The proton acceleration takes place at the "plasma output" boundary, due to these two distinct mechanisms, as soon as the laser pulse and the ultra-relativistic electrons have already left the plasma. We consider such phase of the process, for the proton and heavier species, in a fluid-like description.

In fact, since the laser pulse represents the only "driver" moving the system (the plasma under examination) out from the equilibrium and able to generate strictly kinetic effects on the electrons (as for example the wave-breaking mechanism), once the ultrashort laser pulse has left the plasma and the ultra-relativistic electrons travelling approximately at the speed of the light do the same, the plasma remain without "external force".

As a consequence, the protons and the heavier species dynamics can be considered as a relaxation phase of the plasma, not anymore subjected to the strong external laser field and free from strictly kinetic effects.

For this reason, the physical mechanisms of acceleration we are discussing can be qualitatively illustrated by considering a neutral fluid element consisting of 3 fluid species; the electrons, the protons and the heavier ions considered, with the respective numerical densities $n_e$, $n_p$, $n_{Atom}$, and accelerations $a_e$, $a_p$, $a_{Atom}$. The suffix "Atom" represent the atomic species in the gas-mixture with mass number A and atomic number Z: in the plasma we consider it a fully ionized ion.

Additionally, since the protons and the heavier species due to their higher inertia move on a longer time scale with respect to the electrons taking part in the magnetic vortex structure, the induced electric field can be considered for their dynamics as an external field.

We will consider in the next section the acceleration ratios due to respectively the electrostatic effect owing to the space charge separation, and, the effect of the induced electric field owing to magnetic vortex.

*5.1 Acceleration ratio from space charge effect*

Assuming the Coulombian repulsion (in the ideal conditions of *fully ionized plasma* in which all the electrons are suddenly removed), it follows that the accelerating force applied to the two remaining species, namely protons and ions, within the same fluid element, will be subjected to the same applied force density due to the action-reaction principle. This leads to the estimate of ratio between acceleration of the two species in the fluid elements as:

$$\frac{a_p}{a_{Atoms}} = \frac{n_{Atoms}}{n_p} A, \tag{1}$$

where A is the mass number of the specific atomic species under consideration. As it is shown, Eq.1 clearly introduces two distinct ways of promoting the proton acceleration, a first one through the use of heavier atoms (A>>1) and the other, by changing the species densities. From this simple description an additional generalization to promote the acceleration of heavier ions can be made by simply considering a second atomic species instead of protons, following the formula:

$$\frac{a_{Atoms1}}{a_{Atoms2}} = \frac{n_{Atoms2}}{n_{Atoms1}} \frac{A_2}{A_1}. \tag{2}$$

Eq.2, likewise the proton case is consistent with Eq.1, and also clearly shows two distinct ways to promote the acceleration of the atomic species referred as "Atoms-2" with respect to the atomic species "Atoms-1". This corresponds to either choose different mass number in order to modify their ratio, or to change the densities of the species in order to promote one of them.

These equations are useful to see the scaling, in this mechanism, of the acceleration ratio with dimensionless parameters as the densities ratio or the mass numbers. Nevertheless, these equations only offer a qualitative indication. This means that such equations cannot be used as predictive of the final particles energy. Moreover, since the simulations are stopped after 1ps from the beginning of the interaction when the acceleration mechanism is still acting, it is clear that the final particle energy in experiments will be higher. As an example, considering Eq.1 and the simulation described in Figure 4: Eq.1 clearly give an estimate for the acceleration ratio of more than 2 thousand times the case of pure hydrogen, on the contrary only a factor of 2 is evident from the simulation due to the limited acceleration time of 1ps. For this reasons these equations are only useful to compare the scaling of the mechanisms considered.

As we previously mentioned, it was assumed that such description is strictly ideal and, as a consequence before reaching the stage for a practical implementation, other aspects, such as for example the role of a realistic input density profile, the different ionization potentials of different species and others, have unquestionably to be discussed.

*5.2 Acceleration ratio from external fields*

This case is particularly consistent with the formation of the fields due to the fast electrons escaping from the plasma and generating a magnetic field that in turn induces an electric field, the so called magnetic vortex acceleration [8]: i.e., an induced electric field that further pushes the protons. We consider the ideal simplified case in which, since the fields are generated by the electrons, the ions feel such field as an external one.

In this case we can rewrite the equations for the acceleration of protons and heavier ions, thus yielding:

$$\frac{a_p}{a_{Atoms}} = \frac{Z}{A} \tag{3}$$

Taking into account the arrangement of the terms of Eq.3, the relation is clear: once the atomic species have been selected, Eq.3 gives a fixed value of accelerations due to the declined ratio of densities.

A similar generalization can be done for the case of two distinct heavier atomic species, as follows:

$$\frac{a_{Atoms1}}{a_{Atoms2}} = \frac{Z_1}{A_1} \frac{A_2}{Z_2}. \tag{4}$$

Eq.4, in close likeness to the proton case, shows that also using heavier species the ratio of the accelerations is fixed solely by the atomic and mass numbers.

These equations are useful to see the scaling, in this mechanism, of the acceleration ratio with dimensionless parameters as the mass and atomic numbers. Nevertheless, these equations only offer a qualitative indication. This means that such equations cannot be used as predictive of the final particles energy and are only useful to compare the scaling of the mechanisms considered.

## 6. An ideal case for further studies

The simplified modelling of the previous sections points out the dependence of the scaling of $a_p/a_{Atoms}$ as linked to dimensionless parameters, respectively:

$n_{Atoms}/n_p$, A (eq. 1) and Z/A (eq. 3).

In order to support the interpretation provided above, we consider ideal cases made solely for theoretical considerations. The main idea of this paper can also be used to study the interplay between the magnetic vortex acceleration and the space charge effects since the ratio of the accelerations is different for these mechanisms in the presence of heavier atomic species. This will be the subject of a further study. As extreme examples, made solely for theoretical considerations, we can use a high mass number, for example A=131 of the Xe, and change artificially the Z to verify the effect on the simulations. We present in this section the simulations performed for both an ideal artificial atomic species with Z = 1 and A = 131 (this is the case of Xe$^{+1}$, or ideally a "very heavy proton"), and an ideal artificial species of Z = 131 and A = 131 (or in other words "a fully ionized hydrogen cluster").

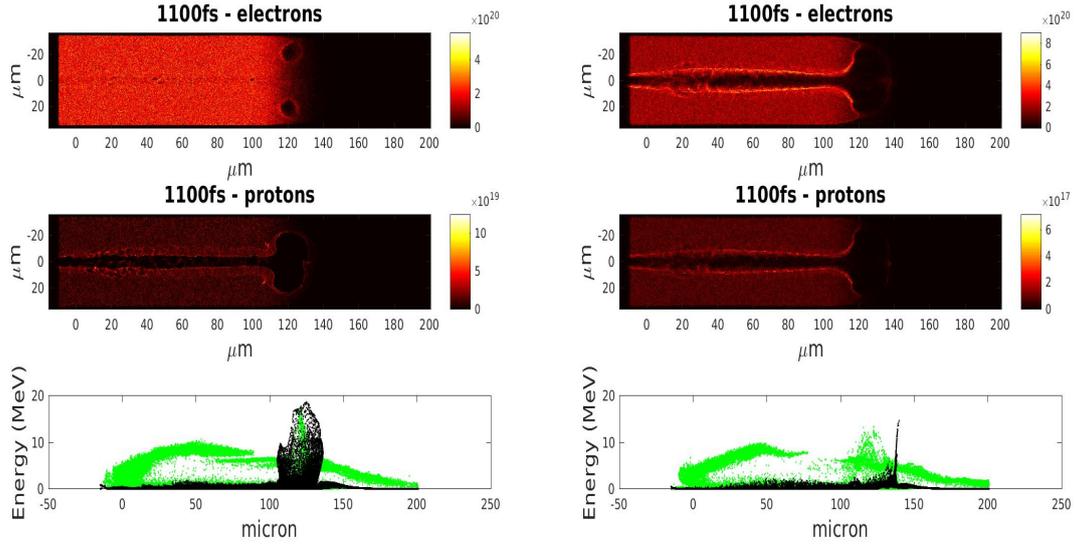

**Figure 6:** shows the snapshots of the laser-plasma interactions, considering a fully-ionized artificial specie with Z=1 and A=131 that we call HeavyProton, in a mixture of 95%**HeavyProton** - 5%**H** (left) and a fully-ionized artificial specie with Z=131 and A=131 that we call ClusterProton, in a mixture of 95%**ClusterProton** - 5%**H** (right) plasma simulated by 2D PIC after 1100fs from the pulse entry (-20 um) in the simulation box (so that the laser pulse has already left the simulated model's area). The total electron density is set to be **$2\times 10^{20}$ e$^-$/cm$^3$**, the plasma profile considered is the same. The electrons (upper row) and protons (middle row) density distributions are shown with the protons (black) and electrons (green) spectra versus the laser propagation direction.

**Figure 6** clearly shows an interesting method to study the interplay between magnetic vortex acceleration and space charge effects: such study has however to take into account the directionality of the accelerated particles if the aim is oriented towards practical applications. From this view point, even if in the maximum of energy the space charge repulsion can offer an advantage due to the possible use of the densities ratio to change the acceleration ratio, it is a non-directional mechanism. On the contrary, magnetic vortex acceleration induces directionality to the accelerated particle bunch. This represents only one of the example that provide an idea of a number of additional features to be considered in further studies towards a possible experimental investigation.

## 7. Conclusions

In this paper we investigated the idea of using gas-mixtures involving high mass number atomic species to promote the proton acceleration in under-dense near critical laser-plasmas. The validity of such idea has been verified with success using 2D-PIC simulations (Epoch). A simple model has been proposed for the main mechanism of the space charge effect, and it was found that the advantage of such mechanism to promote the acceleration is due to the fact that the ratio of the densities of the heavier ions and protons can be tuned to boost the acceleration of the protons.

A basic generalization of such model is presented leading to the possibility of using this concept to additionally accelerate heavier particles as He and C.

Moreover, such method has been shown to give a different ratio for different mechanism in the accelerations of the different species, mainly that of the space charge effects and that of the magnetic vortex acceleration, offering as a consequence viable method for distinguishing these mechanisms in the course of further experimental investigations.

**Author Contributions:** T. Levato had the idea and performed the PIC simulations. L.V. Goncalves contributed to the manuscript. V. Giannini contributed to the physical modelling.

**Funding**: This work has been supported by the project ELI—Extreme Light Infrastructure—phase 2 (CZ.02.1.01/ 0.0/0.0/15_008/0000162) from European Regional Development Fund, by the Ministry of Education, Youth and Sports of the Czech Republic (project No. LQ1606), and by the project "advanced research using high intensity laser produced photons and particles" (CZ.02.1.01/0.0/0.0/16_019/0000789) from European Regional Development Fund.

**Acknowledgments:** The first author thanks Daniele Del Sarto for his invaluable capability in stimulating discussions on the physical mechanisms.